# The combined effects of age and seniority on research performance of full professors[1]

Giovanni Abramo
*Laboratory for Studies of Research and Technology Transfer*
*at the Institute for System Analysis and Computer Science (IASI-CNR)*
*National Research Council of Italy*
ADDRESS: Istituto di Analisi dei Sistemi e Informatica, Consiglio Nazionale delle Ricerche, Via dei Taurini 19, 00185 Roma – ITALY
tel. +39 06 7716417, fax +39 06 7716461, giovanni.abramo@uniroma2.it

Ciriaco Andrea D'Angelo
*University of Rome "Tor Vergata" – Italy and*
*Laboratory for Studies of Research and Technology Transfer (IASI-CNR)*
ADDRESS: Dipartimento di Ingegneria dell'Impresa, Università degli Studi di Roma "Tor Vergata", Via del Politecnico 1, 00133 Roma – ITALY
tel. and fax +39 06 72597362, dangelo@dii.uniroma2.it

Gianluca Murgia
*University of Siena – Italy*
ADDRESS: Dipartimento di Ingegneria dell'Informazione e Science Matematiche, Università degli Studi di Siena, Via Roma 56, 53100 Siena – ITALY
tel. and fax +39 0577 1916386, murgia@dii.unisi.it

**Abstract**

In this work we examine the relationship between research performance, age, and seniority in academic rank of full professors in the Italian academic system. Differently from a large part of the previous literature, our results generally show a negative monotonic relationship between age and research performance, in all the disciplines under analysis. We also highlight a positive relationship between seniority in rank and performance, occurring particularly in certain disciplines. While in Medicine, Biology and Chemistry this result could be explained by the "accumulative advantage" effect, in other disciplines, like Civil engineering and Pedagogy and Psychology, it could be due to the existence of a large performance differential between young and mature researchers, at the moment of the promotion to full professors. These results, witnessed both generally and at the level of the individual disciplines, offer useful insights for policy makers and academia administrators on the role of older professors.

---

# 1. Introduction

In recent decades many of the Western nations' university systems have experienced periods of contraction in faculty recruitment, with the resulting progressive aging of their research personnel (Kyvik and Olsen, 2008; Kyvik, 1990; Bayer and Dutton, 1977). This has stimulated increased interest in investigation of the effects of the advancing age of researchers on their performance (Bonaccorsi and Daraio, 2003; Levin and Stephan, 1989; Cole, 1979). One might expect that the researchers' level of experience could increase with age (Gingras et al., 2008), but in contrast aging could also weigh negatively on the individuals' cognitive capacities (Giambra et al., 1995) and increase the risks for the obsolescence of their knowledge and the loss of interest in research activity (Stephan and Levin, 1992; Price et al., 1975). The empirical studies of the age/performance relationship thus far reported have often furnished contrasting results, which hampers the development of unequivocal indications for policy (Stroebe, 2010).

In this work we analyze the combined effects of age and seniority in rank on the research performance of full professors in Italy, whereby universities are mostly public, competition is scarce and incentive systems insufficient to foster continuous improvements. Our field of observation concerns professors carrying out research in fields where bibliometric techniques can be applied to the measurement of research performance. The dataset, notably broader than previous studies of this kind, consists of 11,989 Italian full professors, who represent 93.6% of the relevant population. The dates of birth of assistant and associate professors were not available to us. Indeed, by focusing on full professors we are able to partly control for the influence of academic rank on level of scientific performance, given that increasing rank corresponds with a general increase in the assets available for research activity, in the form of physical, financial, human or relational resources (Mishra and Smyth, 2013; Abramo et al., 2011; Bozeman and Gaughan, 2011). In addition, concentrating on full professors permits control for the impact of motivational factors exogenous to aging, such as the ambition to attain higher academic rank, which can induce greater productivity (Tien and Blackburn, 1996). The analysis of Italian universities also permits control for economic motivations, since national legislation imposes that all salaries are set according to the same scale of academic rank and seniority, thus making the professors' income completely independent of their individual scientific performance.

Another specificity of the current work is the use, for the measure of research performance, of a sophisticated indicator of research productivity, embedding quantity, impact and relative contribution to output. However the analyses are also conducted for three other indicators previously applied in the literature: one concerning only quantity (total number of the researcher's publications) and two for impact alone – the prestige of the publishing journal as measured by impact factor (IF), and the articles' impact measured by the number of citations received. We adopt a measurement method based on comparative evaluation of performance by academics in the same field and in the same academic rank, starting from publications indexed in the Web of Science (WoS). Each professor is first classified in his or her respective field of research, then the performance is compared with that of the other full professors in the same field, in order to avoid distortions due to the different intensity of publications across fields (Abramo et al., 2008; Butler, 2007; Garfield, 1979). The resort to this classification reduces the direct comparability of our results with those of the preceding literature, but certainly renders them more precise. At the same time, it permits recognition of potential differences across the 11 disciplines of analysis. In fact the relation between age and performance can vary across disciplines, given the different patterns of publication (Bayer and Dutton, 1977; Pfeffer et al., 1976) and the different rates of knowledge obsolescence (Levin and Stephan, 1989; van Heeringen and Dijkwel, 1987). Finally, the different disciplines could register different relations between age and performance due to the effects foreseen under the "accumulative advantage hypothesis" (Allison and Stewart, 1974; Cole and Cole, 1973). According to the hypothesis, the prestige acquired with early career progression should induce the maintenance or increase of the individuals' performance in subsequent years, in part due to the resources obtained through their enhanced reputation (Kyvik, 1990; Cole, 1979). The effects foreseen due to "accumulative advantage" can be different in the diverse disciplines due to the specific manner in which the prestige of researchers is evaluated (Allison and Stewart, 1974), as can likewise vary the capacities to access and exploit various critical resources for research and the successive publication of results (Knorr and Mittermeir, 1980). To evaluate the effects traceable to the accumulative advantage hypothesis in relation to the different disciplines, in this study we will examine the impact on performance, not only of the age of full professors, but also of their seniority in rank.

Because of limited availability of WoS raw data on Italian publications, we adopted a cross-sectional type of analysis, observing the period 2006-2010. Longitudinal analysis seems preferable because in principle it can trace the performance of the individual along his or her whole career, while cross-sectional analysis can only compare different individuals of different ages. However the second can control for the changing research environment which can notably affect performance.

In the following section we discuss the relevant literature on aging and research performance, and in Section 3 we present the methodology for the current study. Section 4 illustrates the results obtained, while in Section 5 we propose future research directions and suggest several policy indications.

## 2. Literature review

Research in the psychological fields has at this point conclusively demonstrated that aging reduces certain cognitive capacities essential to researchers, such as word fluency, attention, visual motor memory, and discrimination (Schaie, 1994). Still, the decline in these capacities only begins in a significant manner at around age 80 (Stroebe, 2010). Since the vast majority of researchers retire before then, biological aging should have very limited impact on performance. Although we note that the large part of outstanding discoveries, as those recognized by the Nobel Prize, are made when the researcher is under age 45 (Dietrich and Srinivasan, 2007; Lehman, 1953), more extensive inquiries have often given contrasting and inconclusive results on the relation between age and research performance. The lack of unanimity in findings is observed both in longitudinal (Kelchtermans and Veugelers, 2011; Gonzalez-Brambila and Veloso, 2007) and in cross-sectional types of analyses. In the cross-sectional analyses, discrepancies are observed both in whether there are one or two peaks in performance as age proceeds (Kyvik, 1990; Bayer and Dutton, 1977) – the relation between age and performance is often non-linear (Levin and Stephan, 1991; Bayer and Dutton, 1977) – and in the age at which these peaks take place (40, 50 or even 60 years) (Gingras et al., 2008; Kyvik, 1990; Stern, 1978). It is notable that due to the limitations in their datasets, these studies investigate the effect of aging in a manner that fails to account for the fact of researchers potentially belonging to different cohorts (Levin and Stephan, 1991; Lawrence and Blackburn, 1988). Researchers from the same cohort share a history of education, values and motivations that may be quite different from those of other cohorts and which may notably affect their levels of productivity (Lawrence and Blackburn, 1985; Bayer and Dutton, 1977). In addition, each cohort of researchers will begin research activity in a given period, characterized by the current level of knowledge matured within the discipline, which in turn influences both the cohort's ongoing levels of productivity and risks of ultimate obsolescence (Levin and Stephan, 1989; Mincer, 1974). The period when a cohort first undertakes its research activity is also characterised by given amounts of available resources, by unique situations in the academic labor market, and even in alternative professions, which can influence levels of individual motivation (Levin and Stephan, 1991). A particular case such as this is that today's younger researchers are being hired and promoted in a period characterized by greater selectivity, compared to previous eras when it was still possible to initiate a faculty career without having matured an international-level scientific curriculum (Kyvik and Olsen, 2008). These factors could offer a partial explanation of why several studies have observed lesser productivity in older cohorts (Levin and Stephan, 1989; Kyvik, 1990). Differently, other cross-sectional studies have given contrary indications, meaning results showing that younger researchers are less productive than older cohorts (Gingras et al., 2008; Bayer and Dutton, 1977). However these studies consider samples of researchers belonging to all academic ranks, which can be a cause of selection bias, since the youngest cohorts will include very large percentages of researchers in early career stages, a number of whom will abandon research altogether as their personal situation becomes more clear (Gingras et al., 2008; Levin and Stephan, 1991; Cole, 1979). Since our research focuses on full professors, who have already matured a level of productivity and are generally unwilling to exit the university system, we propose to validate the following hypothesis:

*H1. In the Italian academic system there exists a monotonic negative relation between the productivity and the age of full professors.*

The conflicting results obtained from cross-sectional and longitudinal analyses have prevented definition of a broadly agreed model for the changes in productivity over researchers' life stages. One of the models having greatest success is that developed by Simonton (1997), which shows how the growth and subsequent decrease in researcher productivity could be due to two opposite effects that vary with career seniority[2]. On the one hand, there is the continuing erosion of the researchers' baggage of innovate ideas, linked to the fact that the academics' rising career seniority will also be accompanied by increasing risks of obsolescence in their knowledge base (Price et al., 1975) and increased difficulty in reorienting towards new research themes (Stephan and Levin, 1992). On the other hand the individual's experience will increase, leading to greater knowledge of their discipline and its typical problems (Gingras et al., 2008), increased effectiveness in direction of laboratory and research teams (Bonaccorsi and Daraio, 2003; Lawrence and Blackburn, 1988), and improved probabilities of obtaining financing (Larivière et al., 2011; Cole, 1992).

---

[2] In general, seniority in career can be considered as a good proxy of age, since the large part of researchers tends to commence their careers at the same age. Various studies show that the two variables (career seniority, age) are highly correlated (Over, 1988; Bayer and Dutton, 1977). Because of this correlation, in this work by "older cohort" we mean both by age and by entry into the academic system.

However, Simonton's model (1997) does not take account of the role of motivation in the productivity levels registered over the researcher's different career stages (Stroebe, 2010). Various economists hold that the observed decreases in productivity are essentially due to the fact that, as the academics' ages increase and their retirement nears, they become less motivated to carry out research, and prefer to devote themselves to other activities that maximize the expected total incomes that they might gain (Kyvik and Olsen, 2008 Diamond, 1984). Still, motivation levels do not depend only on nearing retirement, but also on other factors linked to age in a less direct manner, such as the desire to develop and maintain a professional reputation, which could serve to assist entry into prestigious institutions, or for recognitions such as the keynote speaker role at congresses, or in order to increase the numbers or quality of the professor's students (Stroebe, 2010; Butler and Cantrell, 1989). All these forms of motivation, particularly those independent of external incentivizing mechanisms (Hunter and Kuh, 1987; Finkelstein, 1984), evolve in different ways, but with advancing age tend to increase the differential in productivity between star scientists and other researchers, as foreseen under the accumulative advantage theory (Cole and Cole, 1973). Indeed with increasing age, star scientists register yet more productivity growth, due to the accumulation of resources obtained through previous performance (Cole, 1979; Allison and Stewart, 1974). The resources accumulated can be reputational[3], relational or financial in character, and can support the researcher in accomplishment of more advanced studies, while also inducing continued activity in order to still further increment prestige (Kyvik, 1990). On the other hand, there is a drop in productivity among researchers who reach an intermediate stage in their career and at that point realize they will never achieve notable recognition among their peers (Lawrence and Blackburn, 1985; Allison and Stewart, 1974).

Over time, even academics that previously reached high levels of productivity can experience declining motivation for research activity, for example when possibilities for promotion decrease (Tien, 2000; Kyvik, 1990). In fact various empirical studies document an increase researchers' productivity prior to their promotion, followed by a decrease once the new position is attained (Turner and Mairesse, 2005; Lawrence and Blackburn, 1985). Other studies have demonstrated that once researchers have gone beyond a certain seniority in a given rank they tend to produce less, because they develop the conviction that they will never succeed in reaching any higher rank (Kelchtermans and Veugelers, 2011; Tien and Blackburn, 1996). In this regard, it is notable that star scientists tend to have lesser seniority in rank compared to other scientists (Costas et al., 2010), and that the latter register variation in productivity inverse to their seniority in rank (Kelchtermans and Veugelers, 2011). The impact of promotion and seniority in rank are also observed as being uniquely different for full professors, who having reached the maximum rank can no longer aspire to further promotion. On average, full professors maintain or reinforce their personal productivity even in the first years after promotion (Lawrence and Blackburn, 1985; Blackburn et al., 1978; Katz, 1973), resulting as more productive compared to researchers in other ranks (Abramo et al., 2011; Puuska, 2010; Tien and Blackburn, 1996). In keeping with the

---

[3] The "Matthew effect" itself (Merton, 1968) can be the result of the reputation earned by star scientists.

accumulative advantage hypothesis, the trend in productivity for full professors could be due to the greater accumulation of material and financial resources at their disposal (Mishra and Smyth, 2013; Abramo et al., 2011; Blackburn et al., 1978), as well as their greater capacity to activate collaborations (Abramo et al., 2014b; Bozeman and Gaughan, 2011). Further, full professors are induced to maintain and improve their productivity in order to protect the prestige obtained through their previous performance (Kelchtermans and Veugelers, 2011; Puuska, 2010). Accumulative advantage hypothesis contributes an explanation of why full professors who are nominated at young age, and for years have maintained their reputations and had access to greater resources, obtain levels of post-promotion productivity that are higher than for their colleagues promoted in later age (Turner and Mairesse, 2005). This mechanism is obviously seen primarily in meritocratic systems, such as in highly prestigious universities where promotion is strongly linked to the candidates' demonstrated productivity levels (Finkelstein, 1984; Blackburn et al., 1978). The accumulative advantage hypothesis also contributes to explaining why full professors demonstrate the greatest heterogeneity in levels of post-promotion productivity (Tien and Blackburn, 1996), since this is in turn linked to previous productivity (Hedley, 1987). In fact less productive full professors tend to concentrate on teaching and administrative activity (Hedley, 1987; Stern, 1978), especially when they assume governance positions within their own institutions (Zuckerman and Merton, 1972), because they tend to develop greater affinity with the home institution rather than with their broader discipline (Lawrence and Blackburn, 1985). Differently, star scientists maintain or increase their personal productivity even after promotion to full professor, because rather than depending on further monetary incentives, they seem more motivated by their intrinsic "sacred spark" and the will to preserve their reputation within their scientific community (Kelchtermans and Veugelers, 2011). Beginning from these results as presented in the literature, we propose to validate the following hypothesis:

*H2. In keeping with the accumulative advantage hypothesis, in the Italian academic system there exists a positive relation between productivity of full professors and their seniority in rank.*

In interpreting the empirical evidence on the relations between age, seniority in rank and performance, we should recall that performance has been measured by different bibliometric indicators, ranging from the number of publications, to number of citations, to impact factor of the publishing journals. A number of studies have analyzed indicators of both publication quantity and impact for the same populations, demonstrating different relations with age (Kelchtermans, and Veugelers, 2011; Levin and Stephan, 1989), but the results obtained do not permit definitive conclusions.

The relation between age, seniority in rank and performance might be influenced by such factors as the professor's gender, discipline and university type. The effects of such other factors then need to be controlled for.

The relation between productivity and age could vary according to the discipline considered, because publication patterns vary between disciplines, due to the different working methods adopted (Kyvik, 1990; Bayer and Dutton, 1977), as well to the different orientations of the various academic communities in codifying scientific advancements (Pfeffer et al., 1976). These specificities can determine a different impact of the accumulative advantage effect in the individual disciplines, where it is not only the different manners in which the prestige of the researcher is recognized (Allison and Stewart, 1974), but also the capacity to access and exploit certain resources critical to research activity and the diffusion of its results (Knorr and Mittermeir, 1980). Further, the different relations between productivity and age that are registered between the various disciplines depend also on the different rates of knowledge obsolescence in the various disciplines. In fact some studies (Kyvik, 1990; Levin and Stephan, 1989; van Heeringen and Dijkwel, 1987) have highlighted how in certain disciplines with a rapid rate of obsolescence, such as Physics, Chemistry, Biology and Earth sciences, there is a stronger decrease in productivity with age compared to other less "dynamic" disciplines, such as Economics and Mathematics. However in recent years, the relationship between age and productivity has tended to homogenize across the various disciplines (Costas et al., 2010; Shin and Cummings, 2010; Kyvik and Olsen, 2008).

The gender of researchers can also play a notable role in the relation between productivity and age. In fact women tend to publish less in the first stage of their career (Long, 1992; McDowell, 1982), because of motherhood and the fact that other family responsibilities are often primarily born by women (Larivière et al., 2011; Kyvik and Teigen, 1996). According to some studies (Kyvik and Teigen, 1996), women researchers then subsequently reach a level of productivity very close to that of their male colleagues, while other studies would show that the gap in productivity between the genders remains constant (Larivière et al., 2011). The differences in these analytical results could be due to the fact that the two genders demonstrate very different distributions of productivity. Particularly after 10 years of their career, the least productive women researchers tend to produce more than their least productive male colleagues, while male star scientists continue to produce more compared to women in general (Abramo et al., 2009a; Long, 1992).

The empirical evidence noted, regarding forms of variability between disciplines and genders, stimulates further consideration of the impact of these factors in the analysis of the relations between productivity, age and seniority in rank.

### 3. Data and Methodology

*3.1 Data sources and field of observation*

A brief presentation of several characteristics of the Italian university system assists in interpreting our research results. The Italian Ministry of Education, Universities and Research (MIUR) recognizes a total of 96 universities as having the authority to issue legally-recognized degrees. 94.9% of Italian faculty are employed in the public universities (67). In keeping with the Humboldtian model, there are no 'teaching-only' universities in Italy, as all professors are required to carry out both research and teaching. All new academics enter the university system through national competitions, and career advancement can only proceed by further public competition. The promotions are based on the judgments of committees selected from the existing full professors of the SDSs in question. Differently from associate professors, full professors are eligible to such high level positions as director of schools, departments, PhD programs, etc. They often manage the allocation of resources which affect individuals' research productivity. Moreover as mentor for young researchers, they represent a "role model", influencing their behavior (Pezzoni et al., 2012).

Each academic is classified in one and only one research field, named Scientific Disciplinary Sector (SDS), of which there are 370[4], grouped in 14 disciplines, named University Disciplinary Areas (UDAs). The dataset of Italian professors used in our analysis has been extracted from a database[5] maintained by the MIUR. The database indexes names, academic rank, affiliation, and SDS of all academics in Italian universities. The analysis is limited to the fields where bibliometrics can be applied. Thus we restricted the field of observation to those SDSs (198 in all) where in the five-year period under examination, at least 50% of professors (assistant, associate, and full) achieved at least one publication indexed in the WoS.

The identification of the age and seniority in rank of the full professors was obtained through analysis of lists compiled by the MIUR in 2004, showing the national academics with the right to vote in elections for members in the above-noted career-advancement committees. Since 2008 full professors only have been eligible to be members of career-advancement committees. For this reason, coverage of other academic ranks is insufficient for robust analyses. From these lists we identified the birth dates and dates of appointment to full professor for 11,989 academics, representing 93.6% of the total population of full professors active in the 2006-2010 period, in the 198 SDSs of the field of observation. Table 1 shows that the level of coverage is essentially balanced across the different disciplines, with a minimum of 87% in Earth sciences and a maximum of 98% in Economics and statistics[6]. Missing data refer to professors in very small-sized SDSs, where career advancement competitions have not been launched recently, or to professors who retired before 2010, and were not eligible to vote for evaluation committee members. In general missing data refer mainly to full professors older than the ones analyzed. The average age of full professors as of 31/12/2010 is roughly 58 years in Economics and statistics and over 64 years in Physics. The percentage of inactive full professors varies from slightly over 6% in Physics to over 35% in Economics and statistics. These differences are not necessarily attributable to a lesser emphasis on research on the part of the full professors belonging to any discipline, rather to the diverse patterns of publication that occur. In fact in some UDAs, such as Economics and statistics, Civil engineering, and Pedagogy and psychology, a relevant number of researchers tend to publish exclusively monographs or in national journals, which are not indexed by the WoS.

Figure A1 in SM-Appendix A shows the age distribution for all full professors without distinguishing by discipline. From this we observe that less than 0.5% of full professors are less than age 41 and less than 12% are under age 51. In contrast more than 33% are over 65, with as many as 13% actually being over age 70.

Figure A2 in SM-Appendix A presents the distribution of age at which the researchers in the dataset obtained appointment as full professor. We observe that more than 24% of researchers obtain appointment to this rank before age 41, while just under 10% obtain the appointment after age 55. These results, combined with those on age, suggest that in the majority of cases, the seniority in rank for the full professors is greater than 10 years. On the other hand, the substantial percentage of researchers appointed to full professor at a young age seems to contrast with the results from Pezzoni et al. (2012), who suggest that seniority is an overriding criterion in the promotion of full professors in Italy. Analyzing the age of appointment to full professor by individual UDA, as presented in Table 2, we observe notable differences between the various disciplines. These differences are the fruit not only of different disciplinary publication patterns, but also of the autonomy granted under the Italian system to the individual SDSs for the definition of their relative promotion criteria. While the majority of the full professors in Mathematics and computer sciences are appointed before age 41, in Medicine this occurs in less than 10% of the cases. Economics and statistics, Civil engineering and Industrial and information engineering all show a trend similar to Mathematics and computer sciences, in regards to the percentages of full professors appointed in both young and advanced age. Agricultural and veterinary sciences, while again registering a low percentage of full professors appointed at advanced age, also shows a low percentage appointed at young age. In any case, to evaluate the role of age in the different UDAs, we will verify for the potential presence of a differential in productivity between researchers appointed in the same period, comparing between those appointed at young age and at advanced age.

---

[4] The complete list is accessible on http://attiministeriali.miur.it/UserFiles/115.htm, last accessed January 30, 2015.
[5] http://cercauniversita.cineca.it/php5/docenti/cerca.php, last accessed on January 30, 2015.
[6] The dataset is well-balanced over the individual SDSs given that out of the total 198 SDSs under examination, coverage drops below 67% for only 10 SDSs, where these are also among the smallest of the entire field of observation.

*Table 1: Descriptive statistics of full professors in the dataset, by UDA*

| UDA | Total | Dataset | Coverage (%) | Average age at 31/12/2010 | Average age at appointment to full professor | Inactive (%) |
|---|---|---|---|---|---|---|
| Medicine (MED) | 2,953 | 2,798 | 94.75 | 62.87 | 49.10 | 8.50 |
| Industrial and information engineering (IIE) | 1,981 | 1,837 | 92.73 | 58.83 | 43.64 | 14.08 |
| Biology (BIO) | 1,689 | 1,606 | 95.09 | 62.34 | 45.83 | 7.10 |
| Mathematics and computer sciences (MAT) | 1,120 | 1,090 | 97.32 | 58.51 | 40.57 | 13.75 |
| Chemistry (CHE) | 1,095 | 981 | 89.59 | 63.24 | 46.88 | 4.02 |
| Agricultural and veterinary sciences (AVS) | 1,034 | 928 | 89.75 | 60.63 | 46.27 | 13.44 |
| Physics (PHY) | 918 | 844 | 91.94 | 64.04 | 46.95 | 6.43 |
| Economics and statistics (ECS) | 726 | 714 | 98.35 | 57.97 | 42.60 | 35.40 |
| Civil engineering (CEN) | 560 | 527 | 94.11 | 60.23 | 43.96 | 23.39 |
| Earth sciences (EAR) | 434 | 379 | 87.33 | 63.55 | 47.17 | 14.29 |
| Pedagogy and psychology (PPS) | 305 | 285 | 93.44 | 59.56 | 47.14 | 23.93 |
| Total | 12,815 | 11,989 | 93.55 | 61.26 | 45.78 | 8.11 |

*Table 2: Age distribution at appointment to full professor, by UDA*

| UDA | Full professors appointed before 41 (%) | Full professors appointed after 55 (%) |
|---|---|---|
| Mathematics and computer sciences (MAT) | 58.07 | 4.40 |
| Economics and statistics (ECS) | 42.82 | 5.03 |
| Industrial and information engineering (IIE) | 33.95 | 4.87 |
| Civil engineering (CEN) | 33.78 | 6.26 |
| Biology (BIO) | 22.86 | 9.40 |
| Pedagogy and psychology (PPS) | 21.28 | 12.41 |
| Earth sciences (EAR) | 20.60 | 14.29 |
| Chemistry (CHE) | 18.47 | 14.12 |
| Physics (PHY) | 17.44 | 9.96 |
| Agricultural and veterinary sciences (AVS) | 15.71 | 7.08 |
| Medicine (MED) | 9.47 | 17.26 |
| Total | 25.05 | 10.10 |

*3.2 Performance indicators*

To measure the research performance of the full professors we draw on the Italian Observatory of Public Research (ORP), a database developed and maintained by the authors and derived under license from the WoS. Beginning from the raw data of Italian publications indexed in WoS, and applying a complex algorithm for disambiguation of the true identity of the authors and their institutional affiliations (for details see D'Angelo et al., 2011), each publication[7] is attributed to the university scientists (full, associate and assistant professors) that produced it, with a harmonic average of precision and recall (F-measure) equal to 96 (error of 4%). Beginning from this data we are able to calculate the performance indicators described below, for each full professor. To calculate them one needs to adopt a few simplifications and assumptions. It has been shown (Moed, 2005) that in the hard sciences the prevalent form of codification of research output is publication in scientific journals. The other forms of output which we neglect are often followed by publications that describe their content in the scientific arena, so the analysis of publications alone actually avoids a potential double counting. When measuring labor productivity, if there are differences in the production factors available to each scientist then one should normalize by them. Unfortunately relevant data are not available at individual level in Italy. The first assumption then is that the resources available to professors within the same field of observation are the same. The second assumption is that the hours devoted to research are more or less the same for all professors. Given the main traits of the Italian academic system, the above assumptions appear acceptable.

To ensure satisfactory levels of reliability in the analyses (Abramo et al., 2012a), we measure research performance over a five-year period (2006-2010). We use four indicators of performance.

The first (and most important, in our view) indicator of performance measured is the research productivity, through a proxy called Fractional Scientific Strength (FSS). In formula:

$$FSS = \frac{1}{t} \sum_{i=1}^{N} \frac{c_i}{\bar{c}} \cdot f_i$$

[1]

Where:
t = number of years of work of the professor in the period of observation;
N = number of publications of the professor in the period of observation;
$c_i$ = citations received (at 31/12/2011) by publication *i*;
$\bar{c}$ = average of the distribution of citations received for all cited publications[8] indexed in the same year and subject category of publication *i*;
$f_i$ = fractional contribution of the professor to publication *i*.

Fractional contribution equals the inverse of the number of authors, in those fields where the practice is to place the authors in simple alphabetical order, but assumes different weights in other cases. For the life sciences, widespread practice in Italy and abroad is for the authors to indicate the various contributions to the published research by the order of the names in the byline. For these areas, we give different weights to each co-author according to their order in the byline and the character of the co-authorship (intra-mural or extra-mural). If first and last authors belong to the same university, 40% of citations are attributed to each of them; the remaining 20% are divided among all other authors. If the first two and last two authors belong to different universities, 30% of citations are attributed to first and last authors; 15% of citations are attributed to second and last but one author; the remaining 10% are divided among all others[9]. We do not use a fixed-forward citation window as is customary in studies of this kind, rather we prefer to field-normalize citations by year. Although the evaluation of impact of more recent publications is less robust, in this way we avoid the possible distortions in favor of more recent publications due to the increasing size of the population of potential citing articles.

---

[7] We exclude those document types that cannot be strictly considered as true research products, such as editorial material, conference abstracts, replies to letters, etc.
[8] Abramo et al. (2012b) demonstrate that the average of the distribution of citations received for all cited publications of the same year and subject category is the most effective scaling factor.

Differently from other indicators of research performance, FSS embeds both quantity and impact of production, similarly to the h-index. However, differently from the h-index and most of its variants, it does not neglect the impact of works with a number of citations below h and all citations above h of the h-core works. It does not fail either to field-normalize citations, and to account for the number of co-authors and their order in the byline where appropriate. A thorough explanation of the theory and assumptions underlying FSS can be found in Abramo et al. (2013).

To enable exploration of performance along its individual component dimensions, we also measure an indicator of output alone (P, Publications) and two indicators for impact, with one referring to average impact of the articles (IA, Impact of Articles) and the other to the prestige of the journals in which the professor publishes (IJ, Impact of Journals). In formulae:

$$P = \frac{N}{t}$$

[2]

$$IA = \frac{1}{N}\sum_{i=1}^{N} \frac{c_i}{\bar{c}}$$

[3]

$$IJ = \frac{1}{N}\sum_{i=1}^{N} \frac{if_i}{\overline{if}}$$

[4]

Where:
t, N, $c_i$ and $\bar{c}$ are the same as above;
$if_i$ = impact factor of the journal of publication *i*;
$\overline{if}$ = average of the distribution of impact factors for all journals of publication *i*.

The indicators calculated for each scientist are expressed on a percentile scale (0-100, worst to best) for comparison with the performance of all Italian colleagues (full professors) of the same SDS. This allows to avoid distortions due to the varying intensity of publications across fields (Abramo et al., 2013; Sandström and Sandström, 2009; Zitt et al., 2005).

---

[9] The weighting values were assigned following advice from senior Italian professors in the life sciences. The values could be changed to suit different practices in other national contexts.

To analyze the impact of age and seniority in rank on research performance we assume the follow model of analysis:

*Performance = f(Age, Seniority, Gender, $U_1$, $U_2$, $U_3$)* [5]

where *Performance* is the indicator of research performance, *Age* is the age of the professor measured at 31/12/2010, *Seniority* is the seniority in rank measured at the same moment, *Gender* takes value of 1 if the professor is male and 0 if female. Finally, the variables $U_j$ identify the typology of the professor's home university as of 31/12/2010. The importance of the home university type is shown by preceding studies on the relationship between age and performance (Joy, 2006; Blackburn et al., 1978). In keeping with Tien (2000) and Gonzalez-Brambila and Veloso (2007), we distinguish between private universities ($U_1$) and public ones. In keeping with Shin and Cummings (2010), we also distinguish the universities that are primarily dedicated to research. Although all universities in the Italian system are obligated to conduct research, the Schools for Advanced Studies are unique for their much greater concentration on research, as well as being generally recognized for excellence. Previous research has in fact demonstrated that the average productivity of full professors belonging to these schools ($U_2$) is greater than for the rest of the population (Abramo et al., 2014a). Finally, $U_3$ indicates those full professors belonging to a polytechnic, meaning a university primarily specialized in the engineering disciplines. The analysis of the impact of *Age* and *Seniority* in rank on research performance of full professors is conducted both at the general level and the level of the individual UDAs in order to take account of the impact of the discipline, as discussed above.

Because our performance indicators can be easily transformed into fractional variables, the estimate of the productivity function is obtained by a Quasi-Maximum Likelihood Estimation (QMLE) in the standard logit framework, as previously suggested by Papke and Wooldridge (1996). In accordance with this approach, we estimate the non-linear model:

E(*Performance*|x) = G($\alpha + \sum_{i=1}^{t} \beta_i Age^i$ + $\gamma$ *Seniority* + $\eta$ *Gender* + $\varphi_1 U_1$ + $\varphi_2 U_2$ + $\varphi_3 U_3$) [6]

where G(·) is the logistic function. The choice of a non-linear model is based on numerous studies in the literature (Gonzalez-Brambila and Veloso, 2007; Bayer and Dutton, 1977) that demonstrate a non-linear relation between performance and age. The choice of polynomial degree was made by the minimization of the Akaike information criterion (AIC) (Akaike, 1973), which is one of the most diffused criterion for model selection in Generalized Linear Model (GLM) literature. Through these regressions we obtain coefficients, standard errors and average marginal effects that relate the explanatory variables to the performance indicator under investigation. We can thus evaluate the level of variance (pseudo R-squared) explained by our model (McFadden, 1974). In all the regressions executed we have also checked for the absence of multi-collinearity between the explanatory variables, omitting some of them as necessary. .

Since the propensity of subjects to abandon or retire from university in the period under investigation is independent of their productivity, our analyses are not affected by selection bias. In fact, Italian legislation provides that up to a very advanced age, the choice of retirement for full professors is determined entirely by the individual. Since the maximum threshold for compulsory retirement has changed several times over the past 40 years, we are not able to calculate the exact percentage of full professors that advance their retirement to an age earlier than the cut-off. Still, the fact that there is almost nil correlation between individual performance indicators and the effective retirement age of full professors[10] suggests that research performance is not a determining factor in the choice to remain in the university system.

## 4. Results

To test the two working hypotheses formulated at the outset we first analyze the graphs of the four performance indicators (Section 3.2) related to varying *Age* and *Seniority* in rank. To draw the graphs we regroup the observations in classes of roughly equal numbers, for both *Age* and *Seniority* in rank, and for each class calculate the average percentiles for the performance indicators. Figures 1 and 2 provide graphs showing the trends for the four indicators, by class for *Age* and *Seniority* in rank.

Figure 1 and Figure 2 show that all four performance indicators register a continuous decline with respect to *Age* and *Seniority*. In particular, *P* and *FSS* show marked decreases, while the slopes for *IJ* and *IA* appear slightly less steep. This is due to the fact that the classes of full professors under *Age* 55 obtain better results in *P* and *FSS* compared to the cases for the other two indicators, while the "over-69" class shows inverted results. In the whole population of Italian full professors the performance should decrease more noticeably, as missing data refer mainly to older professors, who retired before 2010 and whose performances (we have measured them) is on average lower than that of our dataset.

However the analysis of these graphs is insufficient to accept or reject the working hypotheses, both because *Age* and *Seniority* in rank are correlated between each other[11] and because the productivity levels also depend on other variables in the model (6). For this reason we conduct a series of QMLE regressions, as presented in Tables 3 to 6. Each table presents the regression results obtained for all four performance indicators (*FSS*, *P*, *IA* and *IJ)*. The tables present the coefficients for the explanatory variables as obtained in the regressions implemented both on the total dataset (Total) and for each discipline[12]. We carried out the regressions at the SDS level in the UDA Physics, and observed that the relation between *Performance*, *Age* and *Seniority* in rank varies to a more or less extent across the SDSs. Because of space constraints in this work we show the results of the analysis at the UDA level. Because we apply the regressions to (almost total) population data, we do not indicate the significance level of the coefficients related to each variable. We report instead their standard errors, which give some descriptive information about the level of heterogeneity of the effects among the individual full professor. Furthermore, we report the average marginal effect of each variable, which is the mean of all marginal effects evaluated for each individual full professor; as for age, the average marginal effects are globally computed by considering all its polynomial degrees. The results concerning the variables proposed in the research hypotheses, age and seniority will be discussed in detail in Sections 4.1 and 4.2; while in the following paragraphs we present the general results and the results concerning the control variables.

---

[10] This correlation ranges from -0.06 (observed for the case *IJ*) to -0.09 (observed for the case *P*).
[11] At the general level, the correlation between *Age* and *Seniority* in rank is 0.7. Breaking down the dataset for professors in the different UDAs, we observe correlation levels between a minimum of 0.66 (in Pedagogy and Psychology) and maximum 0.80 (Economics and statistics). In SM-Appendix B, we provide the complete correlation table for all the UDAs. In any case, as verified under several tests, the high levels of correlation between *Age* and *Seniority* do not cause problems of multi-collinearity in the regressions conducted.

The regressions show a low pseudo R-squared, consistently below 10% for all the performance indicators. There are only a few disciplines, such as Economics and statistics, and Civil engineering, where the variance explained by the statistical model rises above 5% for all the indicators. These results, in line with those from Bayer and Dutton (1977), could be due to the fact that at least some of the processes that influence the relation among performance, age and seniority are state-dependent (Stephan, 1996).

*Figure 1: Graphs of the four performance indicators (percentile) by age class*

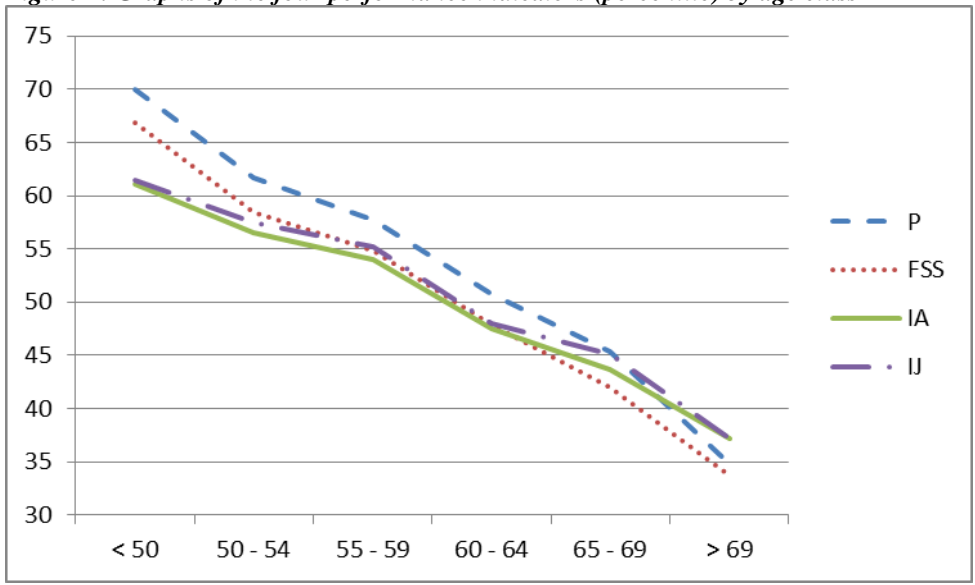

*Figure 2: Graphs of the four performance indicators (percentile) by class for seniority in rank*

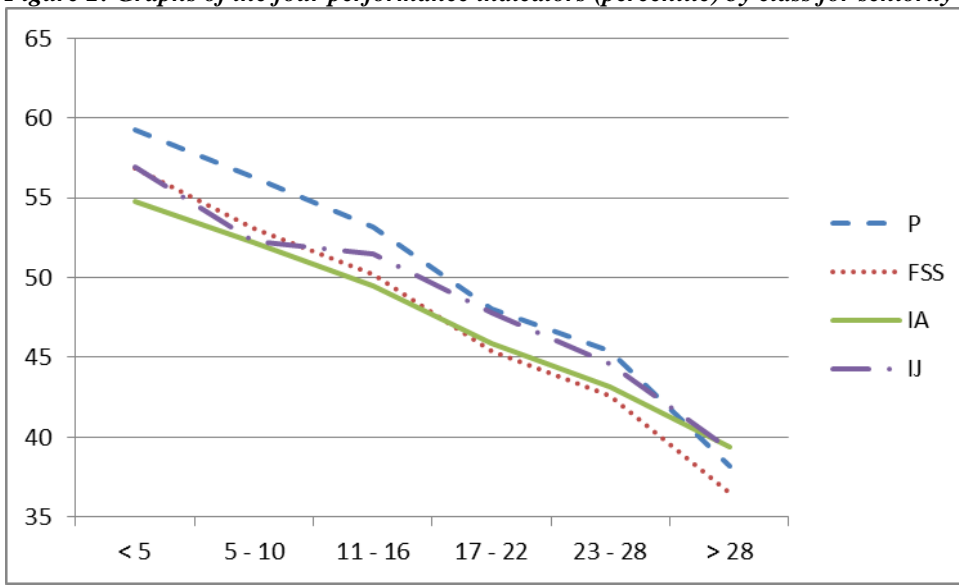

Analyzing the individual explanatory variables, we observe that *male gender* tends to have a positive impact on the four indicators, at both the general level and for the large part of the individual UDAs. Although the standard errors indicate a large heterogeneity in the population, the coefficients seem to show that male professors achieve better research performance. This observation is again in keeping with previous literature (Abramo et al., 2009b; Kyvik and Teigen, 1996; Long, 1992).

For the variables concerning type of university, we observe that belonging to the Schools for Advance Studies has a homogeneous and highly positive impact on almost all the indicators and disciplines. This is due to a few distinctive characteristics of these Schools, which are specifically devoted to highly talented students, with very small faculties and tightly limited enrollment numbers per degree program (Abramo et al., 2014c). However belonging to private university has homogeneous and highly positive impact only in Medicine, a fact that can be traced to the influence of a single university (Milan San Raffaele Vita-Salute University), which features exceptionally high performance in this one discipline. In spite of the focus on engineering, belonging to a polytechnic has a homogeneous and positive impact on only some of the indicators, and primarily in the Biology, Chemistry, and Pedagogy and psychology disciplines.

The choice of the performance indicator seems to have only minimal effect on the relations observed, given that the average marginal effects of age, seniority and some of the other explanatory variables vary in a contained manner between the different indicators. In particular, the effect of seniority and the other control variables seems less dependent on the performance indicator. *Age* seems to have a higher negative effect on *FSS* and *P*, while differences between young and old full professors are less noticeable for the journals' impact factor (*IJ*), and least noticeable for the average citation per article indicator (*IA*). These results differ from those by Levin and Stephan (1989), and by Carayol and Matt (2006), which show a higher negative impact of age on average impact factor than on number of publications. They also differ from those by Diamond (1986), and Kelchtermans and Veugelers (2011) which show a higher negative impact of age on average citations than on number of publications. Stern (1978) and Cole (1979) showed instead that in few disciplines average citations increase with age. Caution is always recommended though when comparing results of studies conducted with different approaches. For example, differently from previous studies, we measure individual performance as compared to that of peers in the same SDS, avoiding the distortions due to different intensity of publication across fields.

*4.1 Impact of age on research performance*

From Table 3, we see that the level of *FSS* almost always tends to decrease with increase in *Age*. In fact there is an inverse tendency only in the case of younger full professors in Chemistry (for *Age* < 44) and Medicine (*Age* < 48). The relation between *FSS* and *Age* is always homogeneous and is generally described by degree 1 or 2 polynomials. Very similar results are also obtained for *P* (Table 4), where *Age* always has a negative effect except for full professors in Chemistry (*Age* < 44) and Medicine (*Age* < 49), for *IA* (Table 5) where *Age* always has negative effect except for full professors in Medicine (*Age* < 45) and for *IJ* (Table 6) where *Age* always has negative effect except for full professors in Industrial and information engineering (*Age* < 43) and in Medicine (*Age* < 43). Other than partially for some UDAs such as Medicine, Civil Engineering, and Chemistry, the negative impact of *Age* on research performance seems confirmed, in keeping with hypothesis *H1*. A further confirmation of hypothesis *H1* is obtained through the analysis of the *Age* of inactive full professors, which is almost always over 54 years, and in a quite homogenous manner among the different UDAs.

The confirmation of hypothesis *H1* also appears in line with previous cross-sectional analyses published in the literature, which show that in disciplines such as Biochemistry (Levin and Stephan, 1989), Chemistry (Cole, 1979), Psychology (Over, 1982) and Medicine (Shin and Cummings, 2010), the number of articles published over a given period of analysis tends to diminish in a trend similar to that shown in our own analyses, especially as concerns *P*. The results for the four indicators in Physics give indications similar to those recorded by Levin and Stephan (1989) in the same discipline, for the indicators of impact and quantity of publications. Beyond these specific cases of agreement with our analyses, the results of the studies cited and from other cross-sectional studies (Kyvik, 1990; Over, 1988; Stern, 1978) are different from ours, above all because at times they show a lower performance for the range of lower ages. Rather than reflecting on the context analyzed, this difference is probably due to the fact that the samples used in the above studies embed a selection bias. These samples are characterized by the presence of an important percentage of young researchers who will likely abandon research precisely because they are unproductive, while in our analysis this problem is overcome by the focus on full professors only. In this case, the negative impact of *Age* on research performance could be attributed either to aging and changing personal interest or to the behavior of the cohort to which the full professor belongs, but, given the cross-sectional nature of our analysis, we are unable to achieve certainty in attributing this negative impact to either causes. If the hypothesis of lesser productivity in the older cohort were correct, such a phenomenon could be due to the greater risk of knowledge obsolescence experienced by the older cohort, or to the ever increasing promotion based on publications at the international level. The first explanation does not seem confirmed by the analysis of the relation between *Age* and performance in the individual UDAs. In fact analyzing the marginal effect of *Age* on the four indicators, the UDAs that show greater average marginal decrease are Civil engineering and Pedagogy and psychology. To our knowledge, these two disciplines seem less affected by obsolescence in comparison to others, such as Physics or Biology. The SDS classification of Italian professors allows for a fine-grained analysis at the SDS level within each UDA. Because of space constraints we report here the example of Physics, in particular those SDSs where the number of professors is large enough to allow for significant results (Table 7[13]). The maximum negative impact of age on performance occurs in FIS/02 (Theoretical Physics, Mathematical Models and Methods). The minimum occurs instead in such "big science" fields as FIS/01 (Experimental Physics), FIS/04 (Nuclear and Subnuclear Physics) and FIS/05 (Astronomy and Astrophysics). The different collaboration behavior and co-authoring practices of scientists in the respective SDSs could in part explain the findings. In Experimental Physics, Nuclear and Subnuclear Physics, and Astronomy and Astrophysics, research projects generally see the collaboration of a number of research labs, which shows in the very high number of co-authors listed in the byline. The contribution to the project is generally intended as of the whole lab, independently of the varying contributions of single members. In such contexts, the impact of age is lessened. Differently, in Theoretical Physics, Mathematical Models and Methods, where the collaboration among labs is less stringent as a requisite to produce results, which shows in shorter authors' lists, the impact of age on performance is more evident.

---

[13] The varying relation between age and performance across SDSs cannot be attributed to the conversion of the dependent variable FSS into percentile. As shown in Table 7, the coefficients of variation in each SDS are very close.

Even though the cross-sectional nature of our analysis does not allow for a definitive conclusion, the analysis of the impact of *Seniority*, presented in the next section, serves in part to evaluate whether the lesser productivity of older cohorts is in fact due to the greater importance given to international scientific publication for promotion to full professor.

*Table 3: QMLE results for each UDA, dependent variable = FSS*

| FSS | Total | CEN | IIE | AVS | BIO | CHE | EAR | ECS | PHY | MAT | MED | PPS |
|---|---|---|---|---|---|---|---|---|---|---|---|---|
| Intercept | 1,087.551 (512.236) | 588.724 (91.714) | 376.053 (45.404) | 402.857 (71.067) | 462.909 (58.08) | -383.917 (373.461) | 490.847 (121.096) | 358.431 (70.743) | 416.286 (88.157) | 319.192 (56.668) | -521.928 (243.27) | 475.5 (122.591) |
| Age | -48.761 (26.169) [-1.451] | -11.344 (1.814) [-2.499] | -6.577 (0.908) [-1.522] | -6.99 (1.36) [-1.658] | -7.932 (1.072) [-1.879] | 19.531 (12.414) [-2.134] | -8.067 (2.053) [-1.89] | -7.016 (1.448) [-1.47] | -6.68 (1.516) [-1.588] | -5.746 (1.09) [-1.337] | 22.327 (7.912) [-2.269] | -9.267 [-2.109] |
| Age² | 0.789 (0.441) | – | – | – | – | – | – | – | – | – | – | – |
| Age³ | -0.005 (0.002) | – | – | – | – | -0.225 (0.104) | – | – | – | – | -0.235 (0.065) | – |
| Seniority | 0.632 (0.285) [0.151] | 4.421 (1.531) [0.974] | 0.073 (0.813) [0.017] | 1.383 (1.192) [0.328] | 2.157 (0.798) [0.511] | 3.882 (1.074) [0.925] | 1.408 (1.471) [0.331] | -0.986 (1.392) [-0.207] | 0.7 (1.105) [0.166] | -0.345 (0.9) [-0.081] | 2.635 (0.644) [0.631] | 3.975 (1.95) [0.905] |
| Gender | 12.958 (5.425) [3.089] | 59.772 (43.732) [13.169] | -5.677 (21.617) [-1.314] | 0.596 (20.102) [0.141] | 14.635 (11.767) [3.467] | -17.168 (19.155) [-4.09] | 2.695 (33.444) [0.632] | 8.956 (21.186) [1.876] | -4.366 (26.631) [-1.038] | 16.757 (16.925) [3.91] | 23.686 (12.403) [5.662] | 7.183 (25.948) [1.635] |
| Polytechnic | -11.635 (8.24) [-2.774] | -4.491 (23.023) [-0.991] | 5.717 (11.523) [1.323] | – | 209.559 (214.53) [49.638] | 101.391 (67.452) [24.156] | -49.427 (74.465) [-11.581] | 9.256 (75.806) [1.939] | 33.723 (39.12) [8.016] | -7.831 (26.677) [-1.823] | – | 123.286 (257.04) [28.064] |
| Private | 29.631 (12.555) [7.064] | – | 10.658 (69.323) [2.467] | 28.781 (59.79) [6.826] | 48.701 (39.902) [11.536] | -94.251 (204.276) [-22.455] | – | 11.06 (25.805) [2.317] | 47.836 (103.733) [11.37] | -61.732 (70.699) [-14.367] | 101.628 (21.612) [24.292] | -46.613 (50.91) [-10.61] |
| "Advanced Studies" | 112.622 (27.547) [26.849] | 83.536 (177.611) [18.405] | 63.078 (76.706) [14.602] | 227.637 (179.988) [53.992] | 149.986 (85.67) [35.527] | 290.207 (417.788) [69.139] | – | 184.188 (109.479) [38.583] | 71.227 (50.542) [16.93] | 133.875 (55.908) [31.158] | 234.132 (182.519) [55.964] | 232.986 (152.567) [53.035] |
| N | 11,989 | 527 | 1,837 | 928 | 1,606 | 981 | 379 | 714 | 844 | 1,090 | 2,798 | 285 |
| Pseudo R-squared | 0.0337 | 0.0793 | 0.0538 | 0.0379 | 0.039 | 0.0347 | 0.0466 | 0.089 | 0.0364 | 0.0504 | 0.0325 | 0.0578 |

*The quantities in ( ) are standard errors, while the quantities [·] are the marginal effects*

Table 4: QMLE results for each UDA, dependent variable = P

| P | Total | CEN | IIE | AVS | BIO | CHE | EAR | ECS | PHY | MAT | MED | PPS |
|---|---|---|---|---|---|---|---|---|---|---|---|---|
| Intercept | 1,154.37 (518.748) [-52.791 (26.475) -1.53] | 483.26 (94.115) [-9.019 (1.719) -2.06] | 396.689 (49.567) [-6.051 (0.904) -1.402] | 386.164 (72.632) [-6.354 (1.351) -1.507] | -152.875 (292.743) [11.687 (9.598) -1.937] | -390.869 (376.024) [20.967 (12.511) -2.263] | 493.773 (122.53) [-7.716 (2.07) -1.785] | 400.673 (70.125) [-7.151 (1.414) -1.569] | 468.609 (89.115) [-7.391 (1.53) -1.754] | 305.661 (56.216) [-5.218 (1.078) -1.228] | -571.837 (244.941) [24.351 (7.964) -1.756] | 418.17 (119.252) [-7.684 (2.191) -1.82] |
| Age | 0.882 (0.446) | | | | | | | | | | | |
| Age² | -0.005 (0.002) | - | - | - | -0.159 (0.079) [-1.937] | -0.241 (0.105) [-2.263] | - | - | - | - | -0.253 (0.065) [-1.756] | - |
| Age³ | | - | - | - | - | - | - | - | - | - | - | - |
| Seniority | 0.668 (0.284) [0.159] | 2.16 (1.442) [0.493] | -0.42 (0.803) [-0.097] | 0.8 (1.18) [0.19] | 2.542 (0.821) [0.603] | 4.101 (1.077) [0.971] | 0.609 (1.468) [0.141] | -0.675 (1.32) [-0.148] | 1.093 (1.107) [0.259] | -0.455 (0.889) [-0.107] | 2.228 (0.643) [0.531] | 3.14 (1.889) [0.744] |
| Gender | 12.143 (5.431) [2.886] | 28.814 (42.216) [6.58] | -28.15 (22.215) [-6.52] | 1.777 (20.187) [0.421] | 21.627 (11.737) [5.134] | -17.275 (19.271) [-4.089] | -0.984 (20.307) [-0.228] | 3.967 (20.925) [0.87] | -10.919 (26.749) [-2.592] | 16.283 (16.808) [3.834] | 26.794 (12.413) [6.368] | -4.222 (25.455) [-1] |
| Polytechnic | -9.369 (8.253) [-2.227] | 1.41 (22.481) [0.322] | 0.686 (11.528) [0.159] | - | 207.975 (223.319) [49.367] | 84.811 (66.162) [20.072] | 87.789 (85.582) [20.31] | 14.357 (75.936) [3.15] | 26.874 (38.712) [-1.688] | -12.717 (26.525) [-2.994] | - | -21.088 (201.338) [-4.994] |
| Private | 29.382 (12.685) [6.984] | - | -22.341 (69.648) [-5.175] | 24.238 (59.794) [5.749] | 36.561 (39.579) [8.679] | -137.193 (210.797) [-32.47] | - | 1.187 (25.461) [0.26] | 26.874 (103.622) [6.38] | -42.274 (67.114) [-9.952] | 101.971 (21.966) [24.235] | -29.756 (49.135) [-7.047] |
| "Advanced Studies" | 70.854 (26.004) [16.841] | 100.443 (200.222) [22.937] | 27.785 (75.777) [6.435] | 118.569 (136.732) [28.123] | 97.556 (75.761) [23.157] | 318.375 (499.667) [75.351] | - | 101.665 (107.655) [22.309] | 14.606 (48.003) [3.467] | 101.405 (53.487) [23.873] | 274.29 (214.366) [65.19] | 177.687 (142.219) [42.082] |
| N | 11,989 | 527 | 1,837 | 928 | 1,606 | 981 | 379 | 714 | 844 | 1,090 | 2,798 | 285 |
| Pseudo R-squared | 0.0359 | 0.0647 | 0.0535 | 0.0345 | 0.0367 | 0.0382 | 0.0534 | 0.0869 | 0.0371 | 0.0426 | 0.0361 | 0.0388 |

*The quantities in ( ) are standard errors, while the quantities [.] are the marginal effects*

*Table 5: QMLE results for each UDA, dependent variable = IA*

| IA | Total | CEN | IIE | AVS | BIO | CHE | EAR | ECS | PHY | MAT | MED | PPS |
|---|---|---|---|---|---|---|---|---|---|---|---|---|
| Intercept | 800.759 (488.604) | 3,225.709 (2575.068) | 264.525 (47.782) | 332.456 (71.685) | 294.495 (57.03) | 191.289 (70.024) | 275.323 (114.819) | 326.293 (69.877) | 292.453 (85.979) | 253.757 (55.679) | -260.575 (238.834) | 386.284 (119.038) |
| Age | -36.325 (24.985) [-1.058] | -167.434 (132.586) [-2.343] | -4.704 (0.886) [-1.132] | -5.927 (1.341) [-1.432] | -4.974 (1.036) [-1.208] | -3.249 (1.246) [-0.804] | -4.279 (1.946) [-1.047] | -6.382 (1.429) [-1.365] | -4.872 (1.483) [-1.185] | -4.447 (1.071) [-1.054] | 12.445 (7.755) [-1.277] | -7.624 (2.202) [-1.775] |
| Age² | 0.594 (0.421) | 2.913 (2.251) | - | - | - | - | - | - | - | - | - | - |
| Age³ | -0.004 (0.002) | -0.018 (0.013) | - | - | - | - | - | - | - | - | -0.141 (0.063) | - |
| Seniority | 0.256 (0.282) [0.062] | 3.859 (1.558) [0.875] | 0.288 (0.797) [0.069] | 1.673 (1.18) [0.404] | 0.567 (0.784) [0.138] | 1.486 (0.955) [0.368] | 0.994 (1.443) [0.243] | -0.818 (1.371) [-0.175] | 0.583 (1.093) [0.142] | -0.754 (0.891) [-0.179] | 1.55 (0.633) [0.379] | 3.826 (1.921) [0.891] |
| Gender | 9.423 (5.374) [2.293] | 50.283 (42.83) [11.399] | -0.336 (21.222) [-0.081] | 1.295 (19.946) [0.313] | 8.265 (11.617) [2.006] | -11.279 (18.678) [-2.792] | -16.814 (32.976) [-4.113] | 2.585 (20.967) [0.553] | 9.496 (26.286) [2.309] | 14.492 (16.78) [3.436] | 19.282 (12.304) [4.711] | -1.461 (25.65) [-0.341] |
| Polytechnic | -13.326 (8.158) [-3.242] | -6.486 (22.836) [-1.471] | 1.364 (11.289) [0.328] | - | 142.732 (174.068) [34.647] | 67.68 (62.053) [16.752] | -98.23 (78.012) [-24.031] | 34.1 (74.954) [7.295] | 3.11 (37.983) [0.756] | -16.049 (26.52) [-3.805] | - | 347.871 (619.592) [80.979] |
| Private | 7.72 (12.315) [1.878] | - | -12.065 (68.465) [-2.905] | 50.093 (59.544) [12.103] | 30.406 (38.712) [7.381] | 22.092 (217.413) [5.468] | - | 7.455 (25.583) [1.595] | -48.207 (107.07) [-11.722] | -54.409 (68.972) [-12.91] | 48.427 (19.816) [11.829] | -56.048 (51.068) [-13.047] |
| "Advanced Studies" | 93.163 (26.118) [22.667] | 87.868 (167.229) [19.919] | 30.703 (72.883) [7.391] | 214.413 (173.819) [51.803] | 107.094 (75.388) [25.996] | 89.786 (166.557) [22.224] | - | 179.917 (109.139) [38.488] | 69.094 (49.72) [16.801] | 114.636 (53.095) [27.178] | 119.044 (148.29) [29.079] | 219.439 (150.731) [51.082] |
| Pseudo R-squared | 0.0189 | 0.0612 | 0.026 | 0.0247 | 0.0213 | 0.0072 | 0.0156 | 0.075 | 0.02 | 0.0374 | 0.0167 | 0.0433 |
| N | 11,989 | 527 | 1,837 | 928 | 1,606 | 981 | 379 | 714 | 844 | 1,090 | 2,798 | 285 |

*The quantities in ( ) are standard errors, while the quantities [.] are the marginal effects*

Table 6: QMLE results for each UDA, dependent variable = IJ

| P | Total | CEN | IIE | AVS | BIO | CHE | EAR | ECS | PHY | MAT | MED | PPS |
|---|---|---|---|---|---|---|---|---|---|---|---|---|
| Intercept | 996.021 (491.376) | 3,791.181 (2620.226) | -205.919 (223.903) | 328.323 (71.647) | 336.118 (57.44) | 164.179 (69.919) | 228.692 (114.542) | 389.353 (70.793) | 292.686 (86.244) | 1892.779 (1545.312) | -226.898 (239.473) | 467.404 (122.185) |
| Age | -48.334 (25.139) [-1.119] | -205.715 (135.526) [-2.121] | 10.574 (7.816) [0.635] | -5.668 (1.338) [-1.368] | -5.662 (1.043) [-1.369] | -3.068 (1.246) [-0.759] | -3.23 (1.941) [-0.787] | -7.423 (1.443) [-1.594] | -5.123 (1.489) [-1.24] | -99.377 (81.305) [-1.334] | 11.18 (7.772) [-1.261] | -9.039 (2.258) [-2.082] |
| Age² | 0.83 (0.424) [-1.119] | 3.703 (2.312) | -0.126 (0.069) | - | - | - | - | - | - | 1.793 (1.41) | -0.13 (0.063) | - |
| Age³ | -0.005 (0.002) | -0.023 (0.013) | - | - | - | - | - | - | - | -0.011 (0.008) | - | - |
| Seniority | 0.329 (0.282) [0.08] | 2.77 (1.527) | 0.298 (0.833) [0.072] | 1.224 (1.178) [0.295] | 0.965 (0.786) [0.233] | 1.058 (0.953) [0.262] | -0.713 (1.442) [-0.174] | -0.128 (1.362) [-0.027] | 0.844 (1.096) [0.204] | -0.013 (0.895) [-0.003] | 1.333 (0.632) [0.325] | 5.204 (1.946) |
| Gender | 14.029 (5.377) [3.411] | 66.152 (42.875) [15.156] | 20.998 (21.115) [5.098] | -0.133 (19.966) [-0.032] | 2.366 (11.639) [0.572] | 18.702 (18.674) [4.629] | -7.045 (32.949) [-1.717] | -2.068 (20.987) [-0.444] | 20.777 (26.39) [5.028] | 15.212 (16.852) [3.607] | 28.84 (12.333) [7.031] | -13.992 (25.789) [1.199] |
| Polytechnic | -16.285 (8.158) [-3.958] | -16.245 (22.757) [-3.722] | -2.001 (11.241) [-0.486] | - | 164.84 (185.247) [39.854] | 50.449 (60.949) [12.487] | -70.925 (74.959) [-17.283] | 91.104 (78.963) [19.559] | -2.806 (37.942) [-0.679] | -19.341 (26.403) [-4.586] | 343.27 (619.599) [79.076] |
| Private | 15.83 (12.357) [3.847] | - | 5.928 (67.782) [1.439] | 32.449 (59.317) [7.832] | 14.071 (38.403) [3.402] | 43.145 (38.457) [10.679] | - | 17.403 (25.478) [3.736] | 39.154 (102.379) [9.475] | -46.907 (68.641) [-11.122] | 57.166 (20.107) [13.936] | -49.375 (50.525) |
| "Advanced Studies" | 98.786 (26.605) [24.009] | 41.942 (151.784) [9.609] | -5.831 (71.935) [-1.416] | 240.407 (195.118) [58.026] | 114.163 (76.932) [27.601] | 97.76 (173.694) [24.197] | - | 149.543 (107.755) [32.105] | 125.724 (56.451) [30.425] | 96.152 (52.251) [22.799] | 110.416 (147.568) [26.917] | 324.29 (203.304) [74.704] |
| Pseudo R-squared | 0.0203 | 0.0601 | 0.0207 | 0.0253 | 0.0242 | 0.007 | 0.0182 | 0.086 | 0.0238 | 0.0388 | 0.0567 | 0.0561 |
| N | 11.989 | 527 | 1,837 | 928 | 1,606 | 981 | 379 | 714 | 844 | 1,090 | 2,798 | 285 |

*The quantities in ( ) are standard errors, while the quantities [.] are the marginal effects*

Table 7: QMLE results for significant SDSs of Physics, dependent variable = FSS

| FSS | FIS/01 | FIS/02 | FIS/03 | FIS/04 | FIS/05 | FIS/07 | Total |
|---|---|---|---|---|---|---|---|
| Intercept | 123.126 (140.814) | 444.831 (270.917) | 506.327 (187.708) | 80.474 (536.947) | 119.663 (302.888) | 392.818 (229.825) | 416.286 (88.157) |
| Age | -2.106 (2.469) [-0.515] | -9.034 (4.539) [-2.148] | -8.974 (3.235) [-2.051] | -1.138 (9.259) [-0.276] | -2.256 (4.928) [-0.517] | -6.067 (3.995) [-1.463] | -6.68 (1.516) [-1.588] |
| $Age^2$ | - | - | - | - | - | - | - |
| Seniority | -1.104 (1.828) [-0.271] | 4.556 (3.121) [1.083] | 2.972 (2.497) [0.679] | -2.127 (4.96) [-0.515] | -2.561 (3.646) [-0.587] | 2.297 (3.342) [0.554] | 0.7 (1.105) [0.166] |
| Gender | 29.959 (39.619) [7.329] | 63.708 (97.926) [15.147] | 3.739 (65.614) [0.855] | 41.667 (111.379) [10.088] | 60.648 (139.346) [13.909] | -52.084 (69.186) [-12.564] | -4.366 (26.631) [-1.038] |
| Polytechnic | 24.787 (49.384) [7.329] | -11.509 (121.373) [-2.736] | -47.151 (77.691) [-10.779] | -207.787 (350.688) [-50.306] | - | - | 33.723 (39.12) [8.016] |
| Private | -216.931 (396.949) [-53.806] | - | 257.252 (424.554) [58.806] | - | - | 78.104 (148.176) [18.84] | 47.836 (103.733) [11.37] |
| "Advanced Studies" | 220.999 (212.745) [54.061] | 34.055 (90.915) [8.097] | 153.456 (103.906) [35.079] | 119.47 (240.158) [28.924] | 231.079 (181.611) [52.995] | - | 71.227 (50.542) [16.93] |
| Pseudo R-squared | 0.0161 | 0.033 | 0.065 | 0.024 | 0.064 | 0.026 | 0.0364 |
| N | 334 | 115 | 164 | 57 | 59 | 104 | 844 |
| Coefficient of variation | 1.5466 | 1.4393 | 1.3225 | 1.1162 | 1.6695 | 1.2193 | 1.6008 |

*The quantities in ( ) are standard errors, while the quantities in [ ] are the marginal effects*

## 4.2 Impact of seniority on research performance

As seen in Tables 3 to 6, *Seniority* generally impacts on research performance in a positive manner in all UDAs, with these exceptions: in Economics and statistics and in Mathematics and computer sciences, for all indicators; in Industrial and information engineering, for *P;* and in Earth sciences for *IJ*. These exceptions may be explained by the percentage of professors promoted before aging 41 which, as shown in Table 2, is highest in these UDAs. It seems that in Mathematics and Economics brilliant results are generally reached at young age more than at a later stage. Apart from these UDAs, results on *Seniority* seem to confirm hypothesis *H2* and are thus in line with the cumulative advantage theory, which would indicate that researchers who are named full professor early in their career accumulate greater resources and maintain their reputation longer over time[14]. However unlike those for *Age*, in the majority of case the standard errors of the coefficients for *Seniority* indicate a high heterogeneity in the population, and they always indicate a marginal effect that is less than the corresponding one for *Age*. This result, which limits the explanatory power of the accumulative advantage hypothesis, could be explained by detailing the analyses at the level of the individual UDA[15]. In fact, standard errors of the coefficients for *Seniority* indicate more homogeneity only in Civil engineering, Pedagogy and psychology, Medicine (for all the indicators), in Biology (for *FSS*, *P* and *IJ*) and in Chemistry (for *FSS*, *P* and *IA*). In some of these disciplines, in particular Medicine, Biology and Chemistry, this result could effectively be connected to the advantages anticipated under the accumulative advantage hypothesis. Indeed, it is in disciplines such as these that greater seniority in rank could facilitate the assignment of critical resources, such as the direction of hospital departments or research laboratories, which impact on productivity. For Civil engineering and Pedagogy and psychology, the explanation would seem more complex, although it is interesting to recall, as illustrated in the previous section, that these two UDAs registered maximum negative marginal effects for the *Age* variable. This could indicate, how especially in these two disciplines, there is a gap of productivity among older professors in favor of full professors promoted at a younger age. This gap could have increased along time, in keeping with heterogeneity of full professors' productivity after promotion, as shown in other studies as well (Turner and Mairesse, 2005; Tien and Blackburn, 1996).

---

[14] The use of the seniority in rank of full-professors only, does not allow to account for any accumulative advantage earned beforehand. Unfortunately, we have no data on the seniority in rank as associate

In order to offer a better explanation of the link between research performance and *Seniority* in rank we have measured the different performances at time of promotion by repeating the same regression analysis on a more limited subset of individuals who were more recently appointed to full professor rank, specifically those with *Seniority* less than eight years[16]. In this manner we are able to show and extend what was indicated by Lissoni et al. (2011) and Pezzoni et al. (2012), who demonstrated that age and seniority in rank positively influence the probability of promotion (although not in a linear manner) for Physics researchers in France, and in still greater measure in Italy. In our analyses, since we consider a limited and recent period of appointment, the examination eliminates the effects linked to the changing national evaluation procedures and the increasing emphasis on international publications, as well as the ever-increasing

---

professors. Our variable *Seniority* in rank represents then a proxy for the accumulative advantage of a professor.

[15] To better understand the impact of *Seniority* on the individual performance indicators, we have attempted to codify this variable in a different manner. Using a dummy variable for each year of appointment to full professor, we obtain results that are scarcely high except for certain years, however these do not seem to indicate any recognizable phenomenon in the processes of advancement. Using a dummy variable for the years in which the procedures in the competitions were the same, we do obtain some interesting results at the general level, however these are not at all unequivocal. In fact the analysis evidences that full professors appointed under the competition procedures in effect until 1973 would have better performance in terms of *FSS* and *P*, but worse in terms of *IA* and *IJ*, compared to their same-aged colleagues appointed under more recent procedures. The details of these analyses, which go beyond the scope of the current work, are available from the authors on request.

[16] The full professors with under eight years *Seniority* are the ones who were appointed or confirmed to this rank during the period under examination (2006-2010). Comparing their performance to that of the other full professors in the dataset, the recently-promoted subset shows higher performance under all indicators, in keeping with previous reports in the literature (Turner and Mairesse, 2005; Lawrence and Blackburn, 1985). Verifying such a statement as a proper hypothesis is clearly beyond the scope of the current study and would require more refined analyses, capable of evaluating the impact of other explanatory variables.

heterogeneity of the performance that we would expect under the accumulative advantage hypothesis, particularly after appointment to full professor. For reasons of brevity, Table 8 provides only the results from the regressions relative to *FSS*. Beginning from the coefficients and the average marginal effects obtained from the *Age* variable, it seems evident that the research performance achieved by younger researchers at the moment of appointment to full professor is decidedly higher. Comparing two average researchers, one at age 45 and the other 65, we can readily calculate how the younger researcher has developed a performance that is 25 percentiles higher for *FSS*. This gap results as still stronger in some UDAs, such as Civil engineering, Economics and statistics, and Pedagogy and psychology, where a researcher aged 45 shows a performance that is on average 39, 36 and 35 percentiles higher compared to a colleague aged 65. The fact that the maximum performance gap is registered also in Civil engineering and Pedagogy and psychology, where we have previously noted a maximum negative impact from *Age* and a high positive impact from *Seniority*, seems to advance the hypothesis that the performances measured at promotion to full professor are not homogenous even within the same discipline, but tend to vary with *Age*. In substance, a younger candidate presents a higher level of performance than an older candidate[17]. This could be due to the fact that top performing young candidates climb faster the career ladder, while for the older candidates there is some recognition of the efforts they have made over a greater number of career years, or in areas different from publications in international journals. An alternative explanation could be that in the Italian system, evaluation for promotion considers the individual's scientific production over a longer period than the five years of our analysis[18]. In any case, the presence of standards for promotion that are thus differentiated with respect to age can negatively affect the efficiency of the Italian university system, because they *de facto* render the role of full professor within the system less clear. If the promotion of a young and brilliant researcher can favor the growth of their university and discipline, while such individuals maintain and strengthen their productivity, including through the effects of "accumulated advantage", then the promotion of a researcher at the close of their career gains little for the university system. The phenomena described, linked with other characteristic problems in the career advancement system, such as the alternating policy episodes of freezing and accelerating career advancement (Lissoni et al., 2011; Bonaccorsi and Daraio, 2003) and widespread favoritism (Abramo et al., 2014c; Perotti, 2008), represent deficiencies and challenges still to be met, which hamper the achievement of full efficiency in the Italian university system.

---

[17] Differently from the works by Lissoni et al. (2011) and Pezzoni et al. (2012), our analysis cannot demonstrate if young candidates to full professor positions are requested higher research performance than older ones. Our analysis only shows a gap of performance between young and old full professors at time of promotion.

[18] The fact of applying such strategies would be against the spirit of Italian law on academic promotion, which emphasizes among other considerations the importance of the candidate's most recent scientific production.

*Table 8: QMLE results, for each UDA, relative to FSS, for seniority levels under 8 years*

| FSS | Total | CEN | IIE | AVS | BIO | CHE | EAR | ECS | PHY | MAT | MED | PPS |
|---|---|---|---|---|---|---|---|---|---|---|---|---|
| Intercept | 297.776 (33.495) | 378.291 (170.75) | 297.504 (91.835) | 311.755 (145.109) | 387.456 (110.322) | 346.389 (135.027) | 358.328 (198.469) | 359.622 (122.022) | 424.796 (174.799) | 340.75 (122.169) | -837.887 (531.605) | 390.323 (184.914) |
| Age | -5.377 (0.6) [-1.276] | -8.401 (3.173) [-1.932] | -5.279 (1.678) [-1.205] | -4.885 (2.764) [-1.143] | -7.257 (2.011) [-1.719] | -5.308 (2.465) [-1.267] | -6.134 (3.413) [-1.395] | -7.835 (2.581) [-1.819] | -6.453 (3.049) [-1.47] | -6.521 (2.242) [-1.475] | - | -7.78 (3.15) [-1.762] |
| Age² | - | - | - | - | - | - | - | - | - | - | -0.357 (0.165) | - |
| Seniority | -0.892 (2.552) [-0.212] | 5.875 (12.818) [1.351] | -0.511 (6.915) [-0.117] | -5.506 (9.396) [-1.288] | 5.066 (7.413) [1.21] | -5.486 (9.291) [-1.309] | -4.026 (16.636) [-0.916] | -1.041 (10.969) [-0.242] | -5.073 (11.613) [-1.156] | 1.528 (9.565) [0.346] | -0.536 (4.912) [-0.128] | -2.642 (14.375) [-0.598] |
| Gender | 22.723 (9.788) [5.391] | 33.407 (66.034) [7.683] | 20.256 (38.36) [4.626] | 13.967 (34.907) [3.268] | 10.424 (23.445) [2.469] | -22.536 (34.54) [-5.377] | 61.058 (59.423) [13.884] | 44.517 (33.969) [10.333] | -8.303 (54.58) [-1.892] | 26.846 (39.071) [6.072] | 10.304 (21.237) [2.468] | 45.447 (43.505) [10.295] |
| Polytechnic | 0.701 (17.99) [0.166] | 38.934 (51.265) [8.954] | 1.476 (24.953) [0.337] | - | - | 44.868 (125.063) [10.706] | -24.466 (125.063) [-5.564] | 2.393 (104.613) [0.555] | -29.707 (88.853) [-6.769] | -31.047 (62.69) [-7.022] | - | - |
| Private | 26.426 (22.714) [6.269] | - | 32.573 (223.358) [7.438] | 143.975 (215.962) [33.689] | 19.269 (74.471) [4.563] | -117.896 (205.622) [-28.132] | - | 3.186 (46.405) [0.739] | 41.42 (210.524) [9.438] | -107.229 (209.51) [-24.251] | 60.681 (33.092) [14.532] | 31.481 (96.22) [7.131] |
| "Advanced Studies" | 127.089 (74.576) [30.152] | - | 115.763 (135.977) [26.436] | - | 83.028 (169.33) [19.664] | 240.814 (425.847) [57.462] | - | - | 193.2 (245.284) [44.021] | 42.835 (193.125) [9.688] | 530.715 (731.698) [127.11] | -36.2 (202.601) [-8.21] |
| N | 2,595 | 105 | 393 | 208 | 319 | 182 | 87 | 192 | 122 | 169 | 721 | 97 |
| Pseudo R-squared | 0.0292 | 0.0593 | 0.0235 | 0.021 | 0.0337 | 0.0318 | 0.0473 | 0.0526 | 0.0479 | 0.0529 | 0.0282 | 0.0668 |

*The quantities in ( ) are standard errors, while the quantities [·] are the marginal effects*

# 5. Conclusions

Over recent decades, numerous studies have examined the relation between performance and age, stimulated largely by concerns over the progressive aging of university faculty now observed in many Western nations. In spite of the quantity of studies there are many cases of conflicting results, in part because of the different statistical methodologies applied, the different models for measuring research performance and the researchers' age, and the difficulties in accounting for the roles of other explanatory variables that influence trends in productivity over the various stages of the researcher's life cycle.

The present study contributes to the debate on the relation between research performance and age, through a cross-sectional analysis of full professors' performance in Italy. As compared to previous studies, our work presents several innovative elements, especially at the methodological level. First, in measuring performance we recur to a proxy of productivity which is more appropriate and reliable than indicators applied in prior literature: FSS in fact embeds both quantity and impact of publications and accounts for the contribution of each author to the publication. Second, differently from previous studies, we measure individual productivity as compared to that of peers in the same field, avoiding the distortions due to different intensity of publication across fields. Third, notwithstanding the limits of cross-sectional approaches, thanks to the measurement of relative performances, we are able to control for the changing research environment which can notably affect performance. Fourth, the focus on full professors allowed to control also for factors related to the presence of young assistant/associate professors, which affected several previous studies. Moreover, the focus on full professors should be of special interest for the decision maker, since they are responsible for the management of some critical resources for research and represent a "role model" for young academics (Pezzoni et al., 2012; Sands et al., 1991).

Our analyses show that as age increases there is a high decline in full professors' productivity which is measured, for the first time in studies on the subject, by an indicator that embeds both quantity and impact. This result is also confirmed through application of other performance indicators, such as number of articles and average impact of the individual's publications, and is observed to different degrees in all the disciplines analyzed. Besides, our findings partially show that seniority within the rank of full professor bears positively on the researchers' productivity. This evidence, especially in such disciplines as Biology, Chemistry and Medicine, seems to partially confirm the expectations of the accumulative advantage hypothesis, since productivity tends to be higher among those that have benefitted from more years of the increased resources that derive from full professor status, including accompanying experience and reputation. In substance, an individual who gains promotion to full professor at young age then maintains and increases his/her productivity more than colleagues who are promoted in later age. This mechanism can also be linked to the productivity differential recorded for these researchers at the moment of their promotion. In fact, our analyses show a relevant performance differential, especially in such disciplines as Civil engineering and Pedagogy and psychology, between older and younger researchers promoted in recent years. This differential could in turn depend on promotion mechanisms in the Italian system that tend to favor older researchers, since such a situation then demands less from them in terms of international publications, or at least permits consideration of the production they "accumulate" over an ample time period.

The monotonic negative relation between performance and age shown by our analysis is aligned to the results of other cross-sectional studies (Levin and Stephan, 1989). Differently from some of them, we show that the negative relation holds true for all performance indicators employed. Other cross-sectional studies show a lower performance for the range of lower ages (Kyvik, 1990; Over, 1988; Stern, 1978), which is due to the fact that they observe any academic ranks.

The upside-down U relation between performance and age is typical of findings in longitudinal studies (Gonzalez-Brambila and Veloso, 2007). These studies cover the whole career progress of professors. Because of age, full professors tend to concentrate on the declining part of the upside-down U curve, which confirms the alignment with our results. In accordance with Turner and Mairesse (2005), we show also that post-promotion productivity of full professors, who are nominated at young age, is higher than for their colleagues promoted in later age. This result can be explained by the accumulative advantage hypothesis, but also by a different level of performance at time of promotion to full professor. This phenomenon, detected through the analysis of professors promoted during the five-year period of observation, adds to the possible causes of performance differences between young and old full professors already observed in the literature.

The evaluation of the productivity levels among the various age ranges can help inform the formulation of more adequate policies to manage the current and near-future development of the academic corps in a manner that ensures suitable growth in overall productivity. In the current Italian university system all professors are required to carry out both research and teaching. Each faculty member must provide a minimum of 350 hours per year devoted to education. Salaries are regulated at the central level and are calculated according to academic rank and seniority. None of a professor's salary depends on merit. Moreover, dismissal of unproductive professors is unheard of.

While for assistant and associate professors career progress can have a strong motivational impact on their research performance, for full professors it does not apply. We do not observe, nor are we aware of other studies on the policies and initiatives that universities, in contexts such as the Italian one, have implemented to foster older professors' low productivity or to direct a larger part of their time to activities other than research. We can think of dual contracts for professors, i.e. "research professors" and "teaching professors", and/or monetary incentives based on research, teaching and managerial performances. Of course these schemes must account for the differences between male and female researchers and those belonging to different academic disciplines and types of universities, given that such differences are once again confirmed by the current analyses.

The definition of such innovative policies, especially for a centralized university system as the Italian one, would require further analyses to confirm or correct the conclusions of the proposed study. To assist to that end, we now intend to broaden the temporal horizon of the performance analyses to clarify if the promotion of older researchers is indeed motivated by their observed long-term production. In addition, this type of study would also permit longitudinal-type analyses; permitting better distinction of the effects of the individual's aging from other characteristic effects linked to his or her cohort of peers. In this way we could overcome one of the main limits of this work, which, being cross-sectional, is not able to disentangle age and cohort effects (Stephan, 1996). Another useful broadening of the analysis would be to examine the academics of other ranks (assistant, associate professors), birthdates allotting, to have a complete picture of the relations between productivity and age, also taking into account the bearing of promotion expectations on researcher performance. Finally, the availability of information on the date of promotion to higher academic ranks, would allow to better account for the accumulative advantage effect,